# The Neutron Decay Into a Proton and an Electron Faults the General Quark Hypothesis

*Roger Ellman*

Abstract

It is well known and well established by scientific observation that a free neutron radioactively decays into a proton plus an electron plus an anti-neutrino with a mean life time before decay of about $900$ seconds.

That established fact conflicts sharply with the hypothesis that the neutron is composed of two "down" plus one "up" quark and that the proton is composed of one "down" plus two "up" quarks.

That conflict throws doubt on the entire quark hypothesis.

Roger Ellman, The-Origin Foundation, Inc.
http://www.The-Origin.org
320 Gemma Circle, Santa Rosa, CA 95404, USA
RogerEllman@The-Origin.org

# The Neutron Decay Into a Proton and an Electron Faults the General Quark Hypothesis

## Roger Ellman

Physicists hypothesize that the fundamental atomic particles, protons and neutrons [and their anti-particles] are composed of Quarks [and their anti-quarks]. In particular the proton is hypothesized to be the combination of two "up" quarks and one "down" quark and the neutron to be the combination of one "up" quark and two "down" quarks. The charges attributed to those quarks are as follows.

| Quark | Charge |
|-------|--------|
| Up    | $+2/3$ that of electron, q |
| Down  | $-1/3$ that of electron, q |

*Table 1*

On that basis the hypothesized quark components produce the verified atomic particle charges as follows.

*(1)*  Proton = U + U + D    Proton Charge = $[+2/3 + 2/3 - 1/3] \times q = +q$
      Neutron = U + D + D    Neutron Charge = $[+2/3 - 1/3 - 1/3] \times q = 0$

However, it is well known and well established by scientific observation that a free neutron radioactively decays into a proton plus an electron plus an anti-neutrino with a mean life time before decay of about *900* seconds as follows.

*(2)*  $^0n_1 \Rightarrow {}^1p_1 + {}^{-1}e_0 + \bar{\nu}$

Of those decay products the proton component requires [per equation *(1)*, above] that one of the neutron's "down" quarks spontaneously change into an "up" quark. That raises the following immediate issues.

  - Which one of the two "down" quarks so changes ?

  - Why not the other one ?

The spontaneous increase in that quark's charge by one full fundamental electric charge, $+q$, is balanced, for the sake of charge conservation, by the electron decay products component with its one full fundamental negative electric charge, $-q$. [The anti-neutrino decay products component is chargeless, its role being only to preserve conservation of energy.] That raises the following immediate issues.

  - The down quark that "decays" into an 'up" quark having a charge of $-1/3$ that of the electron, $q$, how is the charge increase in that quark to $+2/3$ effected ? Where does that one full positive charge, $+q$, come from ?

  - The electron, having a non-fractional charge of $-q$, is not even a quark-related particle; where does it come from in the neutron decay ?

The quark hypothesis for the neutron decay requires one of the neutron's "down" quarks to transform itself into an "up" quark plus an electron and an anti-neutrino. This is described in contemporary physics as a "decay". That raises the following immediate issues.

  - What is meant by the description "decay" for the "down" quarks transformation to "up" ?

- It certainly is not a "radioactive decay" in which an object that is an assembly of fundamental particles emits one or more of those in the process of its decaying into a somewhat lighter / smaller object.

- How can something decay into components that are not part of its pre-decay composition ?

   That quark "down" to "up" transformation requires:

- that it simultaneously increase its charge and decrease its mass [the latter to account for the `neutron - proton` mass differential],

- and do so spontaneously,

- and do so selectively in one of the two neutron "down" quarks while preserving the other unchanged.

That entire set of spontaneous changes would seem quite unlikely. No mechanism is offered and essentially nothing is known of the internal composition and internal behavior of the "down" and "up" quarks.

## *Conclusions*

   - Likely the entire quark hypothesis fails as a description of material reality. Free quarks are not found and the acceptance of the quark hypothesis depends on its specialized construction so as to somewhat or partially systematize particle physics discoveries in a particular fashion.

   - In particular the neutron decay quark transformation hypothesis is much too labored a construct to be seriously entertained.

   - Furthermore, the analysis in "*A New Look at the Neutron and the Lamb Shift*"[1] incontrovertibly demonstrates that the neutron is a proton plus electron combination and that combination is completely compatible with, and to be expected because of, the neutron's natural decay into a proton and an electron without any somewhat fantastic transformation of a quark's charge and mass as the quark hypothesis requires.